# Dual-channel charge transfer doping of graphene by sulfuric acid

Kwanghee Park and Sunmin Ryu*

Department of Chemistry
Pohang University of Science and Technology (POSTECH)
Pohang, Gyeongbuk, 37673, Republic of Korea
E-mail: sunryu@postech.ac.kr

**Abstract:** Two-dimensional materials represented by graphene and transition metal dichalcogenides undergo charge transfer (CT) processes and become hole-doped in strong mineral acids. Nonetheless, their mechanisms remain unclear or controversial. This work proposes and verifies two distinctive CT channels in sulfuric acids, respectively driven by oxygen reduction reaction involving $O_2/H_2O$ redox couples and reduction of bisulfate or related species. Acid-induced changes in the charge density of graphene were in-situ quantified as a function of oxygen content using Raman spectroscopy. At acid concentrations lower than 6 M, the former channel is operative, requiring dissolved $O_2$. Above 6 M, the degree of CT was even higher because the former is cooperated with by the latter channel, which does not need dissolved oxygen. The mechanism revealed in this study will advance our fundamental understanding of how low-dimensional materials interact with chemical environments.

Surface charge transfer (CT) doping or chemical doping refers to the charge injection by adsorbed dopant molecules (electron donor or acceptor). The method, demonstrated early for conductive polymers[1], is highly effective for low-dimensional materials because of their high fraction of surface atoms. Graphene, a representative two-dimensional (2D) material, strongly interacts with $I_2$[2], $Br_2$[2], $NO_2$[3] and alkali metals[4], and undergoes substantial changes in its charge density and thus Fermi level. Such changes enabled graphene to be used in the single-molecule detection[5], pH sensor[6], photodetector[7], solar cell[8] and so on. Similar chemical doping has also been successfully exploited for semiconducting 2D materials.[9, 10] Despite the wide functional tunability and potential applications enabled by CT doping, however, its mechanistic details still remain unclear or unexplored except for simple dopants that do not involve bond-breaking chemistry during the doping process. For example, single-entity dopants like atomic K[4] and molecular halogens[2] inject charges upon adsorption and remain as monovalent ionic adsorbates. In contrast, the $O_2$-mediated hole doping of 2D materials had been controversial for a considerable period because of the intertwined roles of oxygen, water and substrates.[11, 12, 13, 14] Recent studies showed that $O_2/H_2O$ redox couples drive the oxygen reduction reaction (ORR: $O_2 + 4H^+ + 4e^- \leftrightarrow 2H_2O$) serving as composite hole dopants in graphene and 2D semiconductors.[13, 14]

Although chemical doping with mineral acids has been widely used to enhance electrical conductivity in graphene[6, 15, 16], the CT process itself has not been scrutinized until recently. Because the ORR consumes electrons of graphene efficiently at low pH as confirmed for HCl solution[13], the same process should occur in other acids. However, there may be an additional doping channel where their conjugate bases play a role. For example, chemical doping by $H_2SO_4$ solution may be more complex than that by HCl because of bisulfate ions ($HSO_4^-$) and other potentially redox-active species.[17] Notably, the parent and dissociated forms of sulfuric acids collaboratively induce CT doping in graphite and form intercalation compounds in the form of ($C_{24}^+$ $HSO_4^-$ 2.5 $H_2SO_4$) in the presence of oxidizing agents or electrical activation.[18] Despite the lack of consideration of solvation effect, theoretical calculations predicted sizable CT doping in graphene by $HSO_4$ radicals unlike $H_2SO_4$.[19]

In this work, we report the concentration-dependent dual-channel CT mechanism of graphene in $H_2SO_4$. In-situ Raman spectroscopy was employed with an optical liquid cell to directly monitor the charge density of graphene in a real-time manner. It was revealed that the CT is mainly driven by $O_2/H_2O$ redox couples in the concentration lower than 6 M, and possibly bisulfate species give an additional contribution at a higher concentration.

Single-layer graphene samples were prepared by mechanical exfoliation[11] using kish graphite and Si substrates with 285 nm-thick $SiO_2$ layers.[20] To generate nanopores on the graphene surface, the samples were thermally oxidized at 500 °C for 30 min in a quartz tube furnace containing Ar:$O_2$ gas mixture (flow rate = 200:50 mL/min).[21, 22] AFM (atomic force microscopy) topography was obtained under ambient conditions in a noncontact mode with Si tips with a radius of 8 nm. Raman spectra were obtained using a 514 nm laser with a spectrometer with a spectral resolution of 6.0 cm$^{-1}$.[13, 20] The average power on samples was maintained below 200 μW to avoid potential photo-induced artifacts. In-situ Raman measurements in sulfuric acid solutions were performed using a customized optical liquid cell with a Teflon body and quartz window.[13] The concentration of dissolved $O_2$ was varied by sparging Ar or $O_2$ gas into the solutions at a rate of 250 mL/min.

Figures 1a and 1b show the optical micrograph and AFM height image of a representative nano-perforated sample. The oxidation-induced nanopores were created to facilitate otherwise sluggish molecular diffusion at the graphene-substrate interface and enhance the spatial homogeneity of the CT reaction.[22] Whereas either surface of graphene can accommodate CT, the reaction on the one facing substrates is limited by the interfacial diffusion of redox species.[13, 14] The inhomogeneity-derived G-peak splitting in sulfuric acids could be removed by introducing nanopores.[22] The oxidation procedure typically led to a set of nanopores of ~45 nm in diameter and ~80 μm$^{-2}$ in density as shown in Figure 1b.

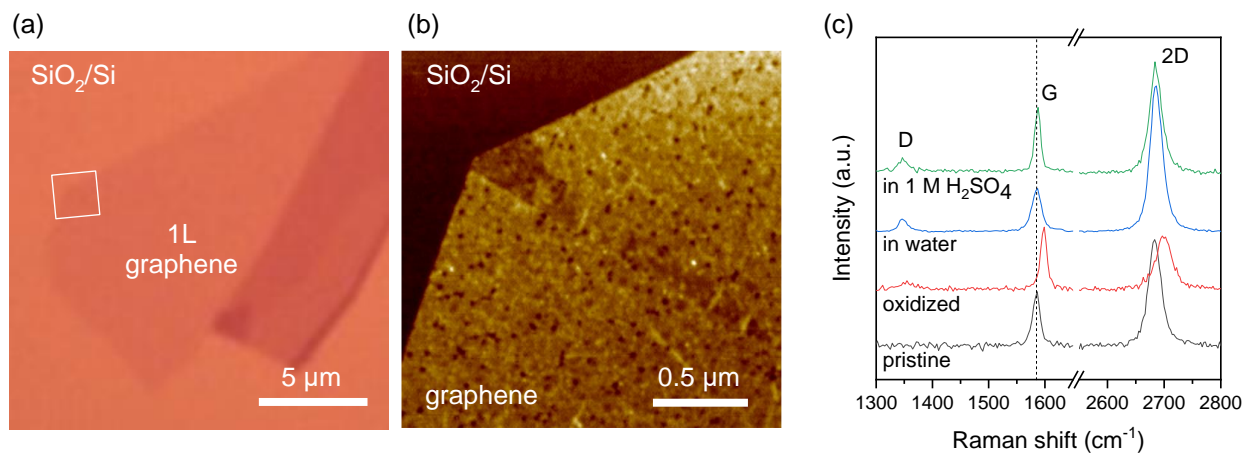

Figure 1. $H_2SO_4$-induced doping in nano-perforated graphene. (a) Optical micrograph of thermally oxidized graphene supported on 285 nm $SiO_2$/Si substrate. (b) AFM height image obtained from the square in (a). (c) Raman spectra of graphene in various conditions.

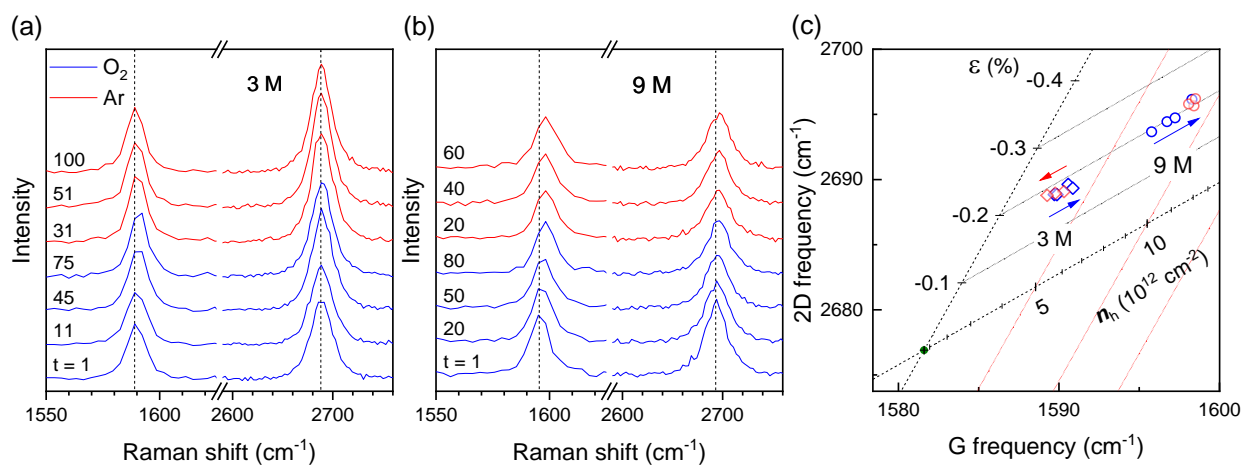

Figure 2. Oxygen-dependent Raman spectra of graphene in $H_2SO_4$. (a & b) Raman spectra of nano-perforated graphene sample in 3 M (a) and 9 M (b) solutions obtained with $O_2$ (blue) or Ar (red) sparged through the solutions. The time (in min) marked for each spectrum was measured since the start of sparging each gas. (c) Vector analysis using 2D & G frequencies from (a) & (b): diamonds for 3 M and circles for 9 M $H_2SO_4$. Blue (red) arrows represent the introduction of $O_2$ (Ar) gas.

Using Raman spectroscopy, we observed the CT doping effect in graphene samples. Figure 1c presents the Raman spectra of the sample in Figure 1a in various conditions. The pristine sample exhibited two Raman peaks, G (~1585 cm$^{-1}$) and 2D (~2680 cm$^{-1}$) that originate from $E_{2g}$ mode at Γ point and overtone of the $A_{1g}$ mode (D peak) near K points, respectively.[23, 24] The lack of D peak (~1350 cm$^{-1}$) activated by structural disorders indicated its high crystallinity, and a high 2D/G intensity ratio suggested negligibly small charge density (< 1×10$^{12}$ cm$^{-2}$).[25] When thermally oxidized, the D peak appeared, the 2D/G ratio decreased, and both of G and 2D peaks upshifted. Whereas the D peak was induced by the edges of the nanopits, the spectral changes of G and 2D peaks can be explained by thermally activated CT doping.[12, 13, 21, 26, 27] Thermal treatments in the range of 200 °C to 700 °C make silica surface more hydrophilic and boost the hole doping of graphene by $O_2$/$H_2O$ redox couples located at the graphene-substrate interface.[13] When immersed in water, the oxidation-induced changes were almost reversed except the D peak. The recovery was made because monolayers of water intercalate the interface and undo the CT reaction.[26] When subsequently immersed in 1 M $H_2SO_4$ solution, the sample again showed the noticeable upshifts of G and 2D peaks (by ~2 cm$^{-1}$), and the 2D/G ratio decreased, which indicates the hole-type CT doping as previously reported.[22, 28]

To verify the role of $O_2$ in the chemical doping by $H_2SO_4$, we obtained the time-lapse Raman spectra in Fig. 2a & 2b while the concentration of dissolved oxygen was varied by sparging Ar or $O_2$ gas into the liquid cell.[13] Before $O_2$ gas was introduced at the time zero, the solution was saturated with Ar gas for 1 hour to minimize the dissolved oxygen. For the 3 M solution in Fig. 2a, G and 2D peaks gradually upshifted until 75 min by 1 cm$^{-1}$ and 0.5 cm$^{-1}$, respectively. Since $O_2$ gas was subsequently replaced with Ar, the two peaks started to downshift and finally recovered their initial positions at 100 min. For the 9 M solution in Fig. 2b, the $O_2$-induced upshifts were larger (~2.5 cm$^{-1}$), which indicated that the degree of CT is even higher for more concentrated solutions. Notably, the upshifts could not be recovered by sparging Ar gas for 60 min. The latter fact suggests another CT route distinctive from the reversible one induced by $O_2$/$H_2O$ redox couples.

In Fig. 2c, we showed how the peak frequencies of G and 2D in Fig. 2a and 2b were varied as the two gases were alternatively introduced. Using the established Raman metrology[27], the G and 2D peak frequencies could be converted into the density of electrical holes ($n_h$) and lattice strain ($\varepsilon$) by projecting the spectral data onto either of the axes representing $n_h$ and $\varepsilon$. It can be readily seen that the overall spectral changes are not due to strain but CT doping for both solutions of 3 M (diamonds) and 9 M (circles). In addition, $O_2$ gas (blue arrows) increased $n_h$ from 3.5 (6.7) to 4.2 (8.5) $\times 10^{12}$ cm$^{-2}$ in the 3 M (9 M) solution. It is also clear that the $O_2$-induced change in $n_h$ was reversed by bubbling Ar (red arrows) at 3 M, but not at 9 M. The reversible change in $n_h$ of 3 M $H_2SO_4$ can be explained by the ORR process, $O_2 + 4H^+ + 4e^-$ (graphene) $\leftrightarrow 2H_2O$[29, 30], and is consistent with the case of HCl solutions.[13] The ORR equation predicts that $n_h$ of graphene should vary reversibly as a function of the oxygen concentration, which explains the change at 3 M, but not 9 M.

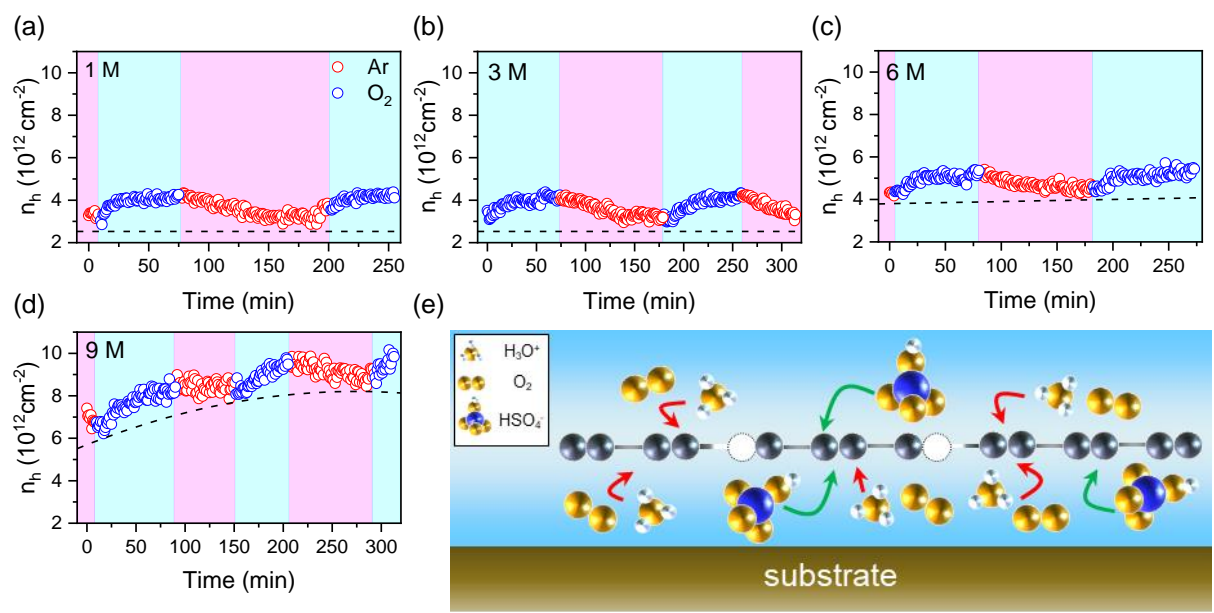

Figure 3. Two competing CT channels. (a ~ d) Time-lapse graphs for $n_h$ of perforated graphene in 1 M (a), 3 M (b), 6 M (c), and 9 M (d) solutions. The data were obtained either in Ar (red) or $O_2$ (blue) gas. The dashed guidelines mark the minimum $n_h$ reached in Ar gas. (e) Scheme depicting a dual-channel CT doping mechanism: ORR by $O_2/H_2O$ redox couples (red arrows) and reduction by bisulfate or related species (green arrows).

In order to elucidate the origin of the spectral irreversibility at high acid concentrations, we obtained Raman spectra in various concentrations. Figures 3a ~ 3d present $n_h$ of a nano-perforated graphene sample in 1, 3, 6 and 9 M solutions during sparging Ar (red) and $O_2$ (blue) gases. In 1, 3 and 6 M solutions, $n_h$ increased during $O_2$ periods from 2.9 to 4.4x10$^{12}$ cm$^{-2}$, 3.0 to 4.3x10$^{12}$ cm$^{-2}$ and 4.3 to 5.3x10$^{12}$ cm$^{-2}$, respectively. The changes in $n_h$ were reversible in multiple Ar-$O_2$ cycles, which indicated that ORR is the main channel for the CT doping. On the other hand, Fig. 3d newly revealed that $n_h$ for 9 M solution is also affected significantly by the gas condition and the apparent irreversibility seen in Fig. 2 is due to the gradual background increase in $n_h$. It is also notable that $n_h$ at the time zero is very high for 6 and 9 M solutions. Most of the acid-induced hole carriers could be reversibly removed when the samples were cleaned with water, which indicated that the spectral changes were not due to photoinduced damage. These facts evidence that hole dopants other than $O_2/H_2O$ couples are operative at elevated concentrations. As proposed for carbon nanotubes[19] and graphite-intercalation compounds[18], bisulfate ions are likely responsible for the slow rise in Fig. 3d. It may be some other species in concentrated sulfuric acids. In Fig. 3e, we depict a scheme for the observed dual-channel CT reactions in sulfuric acids: ORR by $O_2/H_2O$ redox couples (red arrows) and the action of bisulfate ions or related species (green arrow). Below 6 M, the former is the dominant path showing reversible changes in $n_h$, and the latter also acts as an additional doping channel exhibiting much slower kinetics.

In summary, we reported the dual-channel mechanism for the CT doping of graphene in sulfuric acid solutions. Systematic in-situ Raman spectroscopy measurements were performed using an optical liquid cell with control over dissolved gas contents. We revealed that the CT process is driven by the oxygen reduction reaction involving $O_2/H_2O$ redox couples and the reduction of bisulfate ions or related species. The former was the major channel for solutions below 6 M and induced reversible changes in the charge density upon controlling the dissolved oxygen. The latter became an important CT channel for higher concentrations but exhibited no dependence on oxygen gas. The mechanistic details unraveled in the current study will advance our fundamental understanding of how low-dimensional materials interact with various chemical species.

**Acknowledgements**

S.R. acknowledges the financial support from Samsung Research Funding Center of Samsung Electronics under Project Number SSTF-BA1702-08.

**Keywords:** Raman spectroscopy • graphene • charge transfer doping • acid-induced doping • sulfuric acid

**Entry for the Table of Contents**

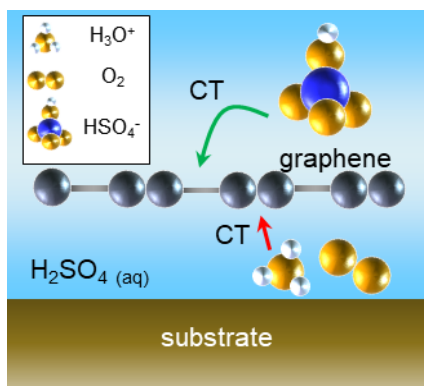

Using Raman spectroscopy, we report two charge transfer (CT) channels of graphene in sulfuric acid: oxygen reduction reaction and reduction by bisulfate ions or related species.